\documentclass[sigconf]{acmart}

\usepackage{balance}

\renewcommand\footnotetextcopyrightpermission[1]{} % removes footnote with conference information in first column.
\begin{document}

\title{THUIR at WSDM Cup 2023 Task 1: Unbiased Learning to Rank}

% 写个5-6页应该就可以了
% 1st. author
\author{Jia Chen}
\affiliation{
    \institution{DCST, Tsinghua University\\Zhongguancun Laboratory}
    \country{Beijing 100084, China}
}
\email{chenjia0831@gmail.com}
% 2nd. author
\author{Haitao Li}
\affiliation{
    \institution{DCST, Tsinghua University\\Zhongguancun Laboratory}
    \country{Beijing 100084, China}
}
\email{liht22@mails.tsinghua.edu.cn}
% 3rd. author
\author{Weihang Su}
\affiliation{
    \institution{DCST, Tsinghua University\\Zhongguancun Laboratory}
    \country{Beijing 100084, China}
}
\email{swh22@mails.tsinghua.edu.cn}
% 4th. author
\author{Qingyao Ai}
\affiliation{
    \institution{DCST, Tsinghua University\\Zhongguancun Laboratory}
    \country{Beijing 100084, China}
}
\email{aiqy@tsinghua.edu.cn}
% 5th. author
%\author{Fan Zhang}
%\affiliation{
%    \institution{Wuhan University}
%    \country{Wuhan, China}
%}
%\email{fan.zhang@whu.edu.cn}
% 6th. author
\author{Yiqun Liu}
\affiliation{
    \institution{DCST, Tsinghua University\\Zhongguancun Laboratory}
    \country{Beijing 100084, China}
}
\email{yiqunliu@tsinghua.edu.cn}

\begin{abstract}
This paper introduces the approaches we have used to participate in the WSDM Cup 2023 Task 1: Unbiased Learning to Rank.
In brief, we have attempted a combination of both traditional IR models and transformer-based cross-encoder architectures.
To further enhance the ranking performance, we also considered a series of features for learning to rank.
As a result, we won 2nd place on the final leaderboard.
%This is an overview of the NTCIR-16 Session Search (SS) task. 
%The task features the Fully Observed Session Search subtask (FOSS) and the Partially Observed Session Search subtask (POSS).
%This year, we received 28 runs from 6 teams in total.
%This paper will describe the task background, data, subtasks, evaluation measures, and the evaluation results, respectively.
%The KUAT team participated in the IR subtask of the NTCIR-16 Data Search 2 Task.
%This paper reports our approach to solving the problem and discusses the official results.
%
%To get familialize the participant papers, please also refer to the past NTCIR proceedings \footnote{\url{https://research.nii.ac.jp/ntcir/workshop/OnlineProceedings15/NTCIR/toc_ntcir.html}}.
%Note that the copyright and page number should be blank in the submitted version; they will be automatically added during the publication process.
\end{abstract}




\maketitle
\pagestyle{plain} % removes running headers for NTCIR proceedings

%\section*{Team Name}
%KUAT

%\section*{Subtasks}
%%Climb the Dubai Tower (Chinese)\\
%%Climb the Dubai Tower (English)\\
%%Climb the Dubai Tower (Japanese)
%IR subtask (English, Japanese)

% introduction of the competition
\section{Introduction}
% ultr的背景
To better organize the result pages in industrial scenarios such as web search engines, e-commerce platforms, and recommendation systems, it is essential to estimate the relevance of a document w.r.t. a specific query or user intent/interest.
With the recent development of deep learning, neural approaches that utilize supervised data (e.g., human annotations) or weak relevance signals (e.g., click labels) to train the ranker have been proposed.
These methods can also be named Learning to Rank (LTR).
Generally, collecting human labels is expensive and labor-intensive.
Although abundant user daily behavioral information such as clicks can be easily collected, directly training the model with user click data may lead to sub-optimal performance because user clicks are usually biased and noisy.
To this end, researchers proposed unbiased learning to rank (ULTR) algorithms that consider multiple biases, e.g., position bias and trust bias, to debias user implicit feedback in an automatic manner. 

% the introduction of the Baidu ultr dataset
Most previous studies evaluated the performance of ULTR methods on synthetic data, which can not intuitively reflect the system's effectiveness as the data distribution in real-world scenarios may be very different from that in synthetic data.
Therefore, the WSDM Cup 2023 task 1 aims at providing a public benchmark dataset for evaluating the performance of various submitted ranking models in the unbiased learning-to-rank setting.
In general, this task provides a large-scale training set that contains billions of web search sessions.
Each session consists of one or more query iteration(s) that record users' interactions with the Baidu search engine within a short time interval.
The training data is organized into multiple fields such as historical query sequences, query reformulation, document rank, title, abstract, vertical type, SERP display time, etc.
Unlike the training data, the validation set is an expert annotation dataset including query, document title, document abstract, query frequency bucket, and five-scale human labels.
However, in Task 1, the annotation data is not allowed for training the ranking model.
The format of the testing set is the same as that of the validating set, while the relevance labels should be further estimated.
More details of the competition dataset can be referred to the home page of this task~\footnote{\url{https://aistudio.baidu.com/aistudio/competition/detail/534/0/introduction}} and the corresponding paper of the Baidu-ULTR dataset~\cite{zou2022large}.
With this setting, the competitors should employ unbiased learning-to-rank techniques to train rankers and estimate the relevance score of each query-document pair in the testing set.

\section{Methodology}
In this section, we will introduce the methods that we have tried to generate the best submission on the leaderboard (finally ranked in the second place), including 1) pre-training and fine-tuning a transformer model, 2) tuning traditional IR models such as BM25 and QL, 3) extracting various features for leaning to rank (LTR).

\begin{table*}[t]
\centering
\vspace{-5mm}
\caption{Features that we used for learning to rank.}
\label{feature}
\begin{tabular}{cll}
\hline
\textbf{Feature ID} & \textbf{Feature Name} & \textbf{Description}\\
\hline
1 & cross\_encoder & Fine-tune the pre-trained transformer model with click data using BCE loss\\
2 & bm25 & BM25 score of title+content using Pyserini (k1=1.6, b=0.87)\\
3 & query\_length & Length of the query\\
4 & title\_length & Length of the title\\
5 & content\_length & Length of the content\\
6 & query\_freq & Frequency bucket of the query\\
7 & ql & Query likelihood score of title+content\\
8 & prox-1 & Averaged proximity score of query terms in title+content\\
9 & prox-2 & Averaged position of query terms appearing in title+content\\
10 & prox-3 & Number of query term pairs appearing in title+content within a distance of 5\\
11 & prox-4 & Number of query term pairs appearing in title+content within a distance of 10\\ 
12 & prox-1-nonstop & PROX-1 score of title+content after being filtered stopwords\\
13 & prox-2-nonstop & PROX-2 score of title+content after being filtered stopwords\\
14 & prox-3-nonstop & PROX-3 score of title+content after being filtered stopwords\\
15 & prox-4-nonstop & PROX-4 score of title+content after being filtered stopwords\\
16 & tf-idf & TF-IDF score of title+content w.r.t. the query\\
17 & tf	& TF score of title+content w.r.t. the query\\
18 & idf & IDF score of title+content\\
19 & bm25\_title	& BM25 score of title using Pyserini (k1=1.6, b=0.87)\\
20 & bm25\_content & BM25 score of content using Pyserini (k1=1.6, b=0.87)\\
21 & bm25-bigram & BM25 score of bigrams in title+content\\
22 & ql-bigram & Query likelihood score of of bigrams in title+content\\
23 & bm25-nonstop & BM25 score of title+content after being filtered stopwords\\
24 & ql-nonstop & Query likelihood score of title+content after being filtered stopwords\\
\hline
\end{tabular}
\vspace{-4mm}
\end{table*}

\subsection{Pre-training and fine-tuning the transformer with the large-scale click data}
Pre-trained language models have shown their effectiveness in re-ranking tasks~\cite{fan2022pre}.
Therefore, we also considered using transformer-based PTMs for better performance in this task.  
To do so, we used the official code released by the organizers as the backbone model and randomized the model parameters without using a warmup checkpoint. 
We pre-train the newly initialized model using a masked language modeling (MLM) loss and a CTR binary cross entropy (CE) loss.
The two losses can be formulated as follows:
\begin{flalign}
	\mathcal{L}_{MLM} = -\sum_{\hat{x} \in m(x)} \mathrm{log}\ p(\hat{x} | x_{\backslash m(x)});\\
	\mathcal{L}_{CE} = -\sum_{c \in \mathcal{C}} c\rm{log}(s) + (1-c)\rm{log}(1-s);
\end{flalign}
where $x$, $m(x)$ and $x_{\backslash m(x)}$ denote the input sequence, the masked and the rest word sets in $x$, respectively.
Besides, $\mathcal{C}$ is the complete click feedback set, and $c$ denotes the click signal on a result, while $s$ is the estimated click probability of this result.

An aggressive mask rate of 40\% is adopted to enhance the representation quality of the language model. 
We stopped the pre-training if the total loss does not decline for 1k steps. Then in the fine-tuning stage, we ignored the MLM loss and merely optimized the CTR loss. 
However, we found that while the CTR loss keeps decreasing after training several epochs, the validating DCG score can not be improved after reaching a threshold of about 7.0. 
At that time, we guessed it may be due to the biases and noises in the click data. 
Up till now, we suspect that there may exist bugs in the PaddlePaddle demo code provided by the official, which is directly transformed from Pytorch with an automatic tool, and the same phenomenon does not appear when training with another Pytorch version of Transformer.

\subsection{Traditional IR methods}
As the click data may be noisy, we also considered employing traditional IR methods to facilitate a robust search.
We utilized the Pyserini~\footnote{\url{https://github.com/castorini/pyserini}} tool to calculate BM25 and query likelihood scores of each query-document pair.
As a result, we found that BM25 with the default parameters yields a good performance of DCG=9.42, which can already rank within top three on the final leaderboard.
We further tuned the parameters of BM25 (k1 and b) on the validation set and achieved a slightly better performance of DCG=9.51 on the testing queries.
The success of BM25 indicates the robustness of these traditional IR methods.
Even in modern search engines, exact matches are still important indicators for relevance estimation.

\subsection{Learning to rank features}
To further improve the ranking performance, we turn to the learning-to-rank (LTR) technique~\cite{yang2022thuir}.
To do so, we have come up with two dozen features to be extracted from the training data.
Besides the scores of traditional IR models, we also considered widely-used LTR features which have been reported useful for improving ranking effectiveness in previous work.
These features include length-related statistics, term proximity-based features, term exact matching features, and their corresponding feature variants, which are calculated based on the raw input after being filtered with stopwords or processed as bigram sequences.
Among them, proximity-based features have been proved effective for improving the performance of a ranker in many previous studies~\cite{chen2022axiomatically, boytsov2011evaluating,fang2004formal,li2023thuir}.
Some other features are inspired by the dataset setting of the TREC Learning to Rank (LETOR) task~\cite{qin2013introducing}.
All the features we have used for competition are listed in Table~\ref{feature}.

\subsection{Other implementation details}
For data preprocessing, we have found that there exist different queries with the same query identifier in the annotation dataset.
To avoid a large discrepancy between our evaluation results and the official results, we remapped the QIDs for all validating queries and ensured that each query only had one unique identifier.
As the size of validation data is quite large, using the whole set for evaluation every time can be time-consuming.
Therefore, we sample a small subset from the validation set according to the maximum semantic similarity between a validating query and all testing queries.
Here we applied a trained PTM checkpoint to generate query vectors and used the [CLS] vector as the embedding of a sentence.
Cosine similarity is used to measure the relevance of two queries.
If the maximum similarity score exceeds a threshold (e.g., top 20\% of all validating queries), we remain the query for validation.

For ensemble approaches, we have considered lightGBM~\cite{ke2017lightgbm} and XGBoost~\cite{chen2016xgboost}. 
We extracted all the features listed in Table~\ref{feature} and employed each ensemble model to estimate the relevance score of each query-document pair.
By using the validation subset we collected aforementioned, we are able to learn the weights of each feature.
We finally employed the trained tree that achieves the highest DCG value on our validation set to infer the relevance score for all testing query-document pairs.

To facilitate the reproducibility of our experiments, we have released our code at this repository~\footnote{\url{https://github.com/xuanyuan14/THUIR_WSDM_Cup}}.

\subsection{An overview of experimental results}
Our experimental results are shown in Table~\ref{performance}.
From the table, we can observe that the best combination of features is using $F_2$-$F_6$, $F_8$-$F_{13}$, and $F_{15}$-$F_{20}$.
Some features are proved useful to improve validating DCG values but can not bring performance gain in the testing queries, such as the cross-encoder score and the query likelihood score.
As mentioned before, there may be problems when training the transformer with the click data.
Merely using the trained transformer for re-ranking yields a very poor performance.
Therefore, it may hurt the system performance on the testing set.
The impact on the validating performance is not evident because the ensemble model may overfit other useful features.

In addition, some features do not work on both the validating and the testing set, e.g., prox-3-nonstop, bm25-nonstop, and ql-nonstop.
As no official stopwords are available, we ranked the tokens in the training corpus with their appearing frequency in descending order and used the top 50 frequent tokens as the stopword set.
Therefore, the selection of the stopword set is not that accurate, which may slightly impact the model performance.
However, we find that term proximity-based features calculated on the sequences after being filtered stopwords can boost the ranking performance by a small margin, indicating that the stopword selection makes sense to some extent.
For the ensemble approach, LightGBM and XGBoost showed close performances when using the same feature subset.
We finally used LightGBM to generate submissions as it achieved a slightly better result on our validating set while using a shorter training time.

\begin{table}[h]
\centering
\caption{Comparison of performances for some methods we have attempted. Concretely, the ensemble model is LightGBM, which performs better than other learning-to-rank methods. The feature IDs can be referred to in Table~\ref{feature}.}
\label{performance}
\begin{tabular}{lccc}
\hline
Method & \# Feature & Dev DCG & Leaderboard \\
\hline
$F_2$(1.2, 0.75) & 1 & 10.12 & 9.42\\
$F_2$(1.6, 0.86) & 1 & 10.3 & 9.51\\
Ensem[$F_1$-$F_{10}$] & 10 & 10.48 & 9.49\\
Ensem[$F_2$-$F_{15}$] & 14 & 10.50 & 9.68\\
Ensem[$F_2$-$F_{13}$, $F_{15}$] & 13 & 10.55 & ? \\	
Ensem[$F_2$-$F_{8}$, $F_{12}$-$F_{13}$, $F_{15}$] & 10 & 10.60 & 9.67\\ 
Ensem[$F_{2}$-$F_{13}$, $F_{15}$-$F_{16}$, $F_{22}$] & 15 & 10.69 & 9.774\\
Ensem[$F_{2}$-$F_{13}$, $F_{15}$-$F_{20}$] & 18 & 10.71 & \textbf{9.91}\\
Ensem[$F_{2}$-$F_{6}$, $F_{8}$-$F_{13}$, $F_{15}$-$F_{20}$] & 17 & \textbf{10.75} & 9.89\\
\hline
\end{tabular}
\vspace{-4mm}
\end{table}

\section{Conclusion}
In this paper, we have briefly introduced the methods we used to participate in the WSDM Cup 2023 Task 1: Unbiased Learning to Rank.
Through the experimental results, we have the following conclusions: 
1) Although transformer-based models have made great success in information retrieval tasks, traditional IR methods such as BM25 can still yield robust performance in real-world web search scenarios.
2) Using some axiomatic features, such as proximity-based features, can further boost the ranking performance.
Based on the techniques in this paper, we have achieved second place on the final leaderboard.
Due to the time limit, we have not tried more sophisticated approaches to debias the click signals with some counterfactual learning to rank algorithms or to effectively train the transformer model yet.
Both perspectives can be regarded as future work.

\section{Acknowledgements}
This work is supported by the Natural Science Foundation of China (Grant No. 61732008) and Tsinghua University Guoqiang Research Institute.

%\newpage
%\balance
\bibliographystyle{ACM-Reference-Format}
\bibliography{ntcirsample}


\begin{thebibliography}{10}

%%% the Web address.

\ifx \showCODEN    \undefined \def \showCODEN     #1{\unskip}     \fi
\ifx \showDOI      \undefined \def \showDOI       #1{#1}\fi
\ifx \showISBNx    \undefined \def \showISBNx     #1{\unskip}     \fi
\ifx \showISBNxiii \undefined \def \showISBNxiii  #1{\unskip}     \fi
\ifx \showISSN     \undefined \def \showISSN      #1{\unskip}     \fi
\ifx \showLCCN     \undefined \def \showLCCN      #1{\unskip}     \fi
\ifx \shownote     \undefined \def \shownote      #1{#1}          \fi
\ifx \showarticletitle \undefined \def \showarticletitle #1{#1}   \fi
\ifx \showURL      \undefined \def \showURL       {\relax}        \fi
% invisible to TeX
\providecommand\bibfield[2]{#2}
\providecommand\bibinfo[2]{#2}
\providecommand\natexlab[1]{#1}
\providecommand\showeprint[2][]{arXiv:#2}

\bibitem[\protect\citeauthoryear{Boytsov and Belova}{Boytsov and
  Belova}{2011}]%
        {boytsov2011evaluating}
\bibfield{author}{\bibinfo{person}{Leonid Boytsov} {and} \bibinfo{person}{Anna
  Belova}.} \bibinfo{year}{2011}\natexlab{}.
\newblock \showarticletitle{Evaluating Learning-to-Rank Methods in the Web
  Track Adhoc Task.}. In \bibinfo{booktitle}{\emph{TREC}}.
\newblock


\bibitem[\protect\citeauthoryear{Chen, Liu, Fang, Mao, Fang, Yang, Xie, Zhang,
  and Ma}{Chen et~al\mbox{.}}{2022}]%
        {chen2022axiomatically}
\bibfield{author}{\bibinfo{person}{Jia Chen}, \bibinfo{person}{Yiqun Liu},
  \bibinfo{person}{Yan Fang}, \bibinfo{person}{Jiaxin Mao},
  \bibinfo{person}{Hui Fang}, \bibinfo{person}{Shenghao Yang},
  \bibinfo{person}{Xiaohui Xie}, \bibinfo{person}{Min Zhang}, {and}
  \bibinfo{person}{Shaoping Ma}.} \bibinfo{year}{2022}\natexlab{}.
\newblock \showarticletitle{Axiomatically Regularized Pre-training for Ad hoc
  Search}. In \bibinfo{booktitle}{\emph{Proceedings of the 45th International
  ACM SIGIR Conference on Research and Development in Information Retrieval}}.
  \bibinfo{pages}{1524--1534}.
\newblock


\bibitem[\protect\citeauthoryear{Chen and Guestrin}{Chen and Guestrin}{2016}]%
        {chen2016xgboost}
\bibfield{author}{\bibinfo{person}{Tianqi Chen} {and} \bibinfo{person}{Carlos
  Guestrin}.} \bibinfo{year}{2016}\natexlab{}.
\newblock \showarticletitle{Xgboost: A scalable tree boosting system}. In
  \bibinfo{booktitle}{\emph{Proceedings of the 22nd acm sigkdd international
  conference on knowledge discovery and data mining}}.
  \bibinfo{pages}{785--794}.
\newblock


\bibitem[\protect\citeauthoryear{Fan, Xie, Cai, Chen, Ma, Li, Zhang, Guo,
  et~al\mbox{.}}{Fan et~al\mbox{.}}{2022}]%
        {fan2022pre}
\bibfield{author}{\bibinfo{person}{Yixing Fan}, \bibinfo{person}{Xiaohui Xie},
  \bibinfo{person}{Yinqiong Cai}, \bibinfo{person}{Jia Chen},
  \bibinfo{person}{Xinyu Ma}, \bibinfo{person}{Xiangsheng Li},
  \bibinfo{person}{Ruqing Zhang}, \bibinfo{person}{Jiafeng Guo},
  {et~al\mbox{.}}} \bibinfo{year}{2022}\natexlab{}.
\newblock \showarticletitle{Pre-training methods in information retrieval}.
\newblock \bibinfo{journal}{\emph{Foundations and Trends{\textregistered} in
  Information Retrieval}} \bibinfo{volume}{16}, \bibinfo{number}{3}
  (\bibinfo{year}{2022}), \bibinfo{pages}{178--317}.
\newblock


\bibitem[\protect\citeauthoryear{Fang, Tao, and Zhai}{Fang
  et~al\mbox{.}}{2004}]%
        {fang2004formal}
\bibfield{author}{\bibinfo{person}{Hui Fang}, \bibinfo{person}{Tao Tao}, {and}
  \bibinfo{person}{ChengXiang Zhai}.} \bibinfo{year}{2004}\natexlab{}.
\newblock \showarticletitle{A formal study of information retrieval
  heuristics}. In \bibinfo{booktitle}{\emph{Proceedings of the 27th annual
  international ACM SIGIR conference on Research and development in information
  retrieval}}. \bibinfo{pages}{49--56}.
\newblock


\bibitem[\protect\citeauthoryear{Ke, Meng, Finley, Wang, Chen, Ma, Ye, and
  Liu}{Ke et~al\mbox{.}}{2017}]%
        {ke2017lightgbm}
\bibfield{author}{\bibinfo{person}{Guolin Ke}, \bibinfo{person}{Qi Meng},
  \bibinfo{person}{Thomas Finley}, \bibinfo{person}{Taifeng Wang},
  \bibinfo{person}{Wei Chen}, \bibinfo{person}{Weidong Ma},
  \bibinfo{person}{Qiwei Ye}, {and} \bibinfo{person}{Tie-Yan Liu}.}
  \bibinfo{year}{2017}\natexlab{}.
\newblock \showarticletitle{Lightgbm: A highly efficient gradient boosting
  decision tree}.
\newblock \bibinfo{journal}{\emph{Advances in neural information processing
  systems}}  \bibinfo{volume}{30} (\bibinfo{year}{2017}).
\newblock


\bibitem[\protect\citeauthoryear{Li, Chen, Su, Ai, and Liu}{Li
  et~al\mbox{.}}{2023}]%
        {li2023thuir}
\bibfield{author}{\bibinfo{person}{Haitao Li}, \bibinfo{person}{Jia Chen},
  \bibinfo{person}{Weihang Su}, \bibinfo{person}{Qingyao Ai}, {and}
  \bibinfo{person}{Yiqun Liu}.} \bibinfo{year}{2023}\natexlab{}.
\newblock \showarticletitle{Towards Better Web Search Performance:
  Pre-training, Fine-tuning and Learning to Rank}.
\newblock \bibinfo{journal}{\emph{WSDM Cup 2023}} (\bibinfo{year}{2023}).
\newblock


\bibitem[\protect\citeauthoryear{Qin and Liu}{Qin and Liu}{2013}]%
        {qin2013introducing}
\bibfield{author}{\bibinfo{person}{Tao Qin} {and} \bibinfo{person}{Tie-Yan
  Liu}.} \bibinfo{year}{2013}\natexlab{}.
\newblock \showarticletitle{Introducing LETOR 4.0 datasets}.
\newblock \bibinfo{journal}{\emph{arXiv preprint arXiv:1306.2597}}
  (\bibinfo{year}{2013}).
\newblock


\bibitem[\protect\citeauthoryear{Yang, Li, Chu, Zhan, Liu, Zhang, and Ma}{Yang
  et~al\mbox{.}}{2022}]%
        {yang2022thuir}
\bibfield{author}{\bibinfo{person}{Shenghao Yang}, \bibinfo{person}{Haitao Li},
  \bibinfo{person}{Zhumin Chu}, \bibinfo{person}{Jingtao Zhan},
  \bibinfo{person}{Yiqun Liu}, \bibinfo{person}{Min Zhang}, {and}
  \bibinfo{person}{Shaoping Ma}.} \bibinfo{year}{2022}\natexlab{}.
\newblock \showarticletitle{THUIR at the NTCIR-16 WWW-4 Task}.
\newblock \bibinfo{journal}{\emph{Proceedings of NTCIR-16. to appear}}
  (\bibinfo{year}{2022}).
\newblock


\bibitem[\protect\citeauthoryear{Zou, Mao, Chu, Tang, Ye, Wang, and Yin}{Zou
  et~al\mbox{.}}{2022}]%
        {zou2022large}
\bibfield{author}{\bibinfo{person}{Lixin Zou}, \bibinfo{person}{Haitao Mao},
  \bibinfo{person}{Xiaokai Chu}, \bibinfo{person}{Jiliang Tang},
  \bibinfo{person}{Wenwen Ye}, \bibinfo{person}{Shuaiqiang Wang}, {and}
  \bibinfo{person}{Dawei Yin}.} \bibinfo{year}{2022}\natexlab{}.
\newblock \showarticletitle{A large scale search dataset for unbiased learning
  to rank}.
\newblock \bibinfo{journal}{\emph{arXiv preprint arXiv:2207.03051}}
  (\bibinfo{year}{2022}).
\newblock


\end{thebibliography}

\end{document}